\documentclass[%
 reprint,
 amsmath,amssymb,
 aps,
]{revtex4-1}

\usepackage{graphicx}
\usepackage{dcolumn}
\usepackage{bm}
\usepackage{hyperref}

\begin{document}

\preprint{APS/123-QED}
\title{Scanning tunnelling microscopy and X-ray photoemission studies of NdNiO$_{2}$ infinite-layer nickelates films}
\author{Martando Rath$^{1}$, Yu Chen$^{1}$, Guillaume Krieger$^{2,3}$, H. Sahib $^{2}$, Daniele Preziosi$^{2}$ and Marco Salluzzo$^{1}$}
 \email{marco.salluzzo@spin.cnr.it}
\affiliation{$^{1}$CNR-SPIN Complesso di Monte S. Angelo, via Cinthia—I-80126 Napoli, Italy}
\affiliation{$^{2}$Université de Strasbourg, CNRS, IPCMS UMR 7504, F-67034 Strasbourg, France}
\altaffiliation[$^{3}$present address:]{Eindhoven University of Technology, P.O. Box 513, 5600 MB Eindhoven, The Netherlands}

\date{\today}

\begin{abstract}
We report scanning tunnelling microscopy (STM) and x-ray photoemission spectroscopy (XPS) measurements on uncapped and SrTiO$_{3}$ (STO) capped NdNiO$_{2}$ realized by pulsed-laser deposition and topotactic reduction process. We find that untreated NdNiO$_{2}$ surfaces are insulating and contain Ni mostly in a nominal Ni$^{2+}$ oxidation state. Room temperature STM shows signatures of a striped-like topographic pattern, possibly compatible with recent reports of ordered oxygen vacancies in uncapped nickelates due to incomplete oxygen de-intercalation of the upper layers. A metallic surface and a full Ni$^{1+}$ oxidation state is recovered by ultra high vacuum annealing at 250~$^{\circ}$C, as shown by STM and XPS. STO-capped NdNiO$_{2}$ films, on the other hand, show Ni mostly in Ni$^{1+}$ configuration, but Nd 3\textit{d}$_{5/2}$ core level spectra have a relevant contribution from ligand 4\textit{f}$^{4}$\underline{L} states, suggesting the formation of a NdTiNiO$_{x}$ layer at the interface with the STO. By \textit{in situ} unit-cell by unit-cell Ar-ion sputtering removal of the STO capping and of the non stoichiometric interface layer, we were able to address the surface electronic properties of these samples, as shown by high resolution valence band photoemission spectroscopy. The results provide insights in the properties of infinite-layer NdNiO$_{2}$ thin films prepared by the CaH$_{2}$ topotactic reduction of perovskite NdNiO$_{3}$ and suggest methods to improve their surface quality.


\end{abstract}

\pacs{Valid PACS appear here}
\maketitle


\section{\label{sec:level1}Introduction}
\par Rare-earth nickelates RNiO$_{3}$ (R= rare-earth) have been the leitmotiv of intense theoretical and experimental research efforts towards the realization of novel high-Tc superconductors mimicking the electronic structure of cuprates, as suggested by a seminal work \cite{Anisimov1999_Per_nickelates}.
The quest for Ni-based superconductivity has eventually led to a successful fabrication of superconducting Nd$_{0.8}$Sr$_{0.2}$NiO$_{2}$ thin films deposited onto (001) SrTiO$_{3}$ (STO) with a T$_{c}$ = 15 K, obtained after topotactic reduction of perovskite NdNiO$_{3}$ grown by pulsed laser deposition (PLD) \cite{Li2019_SC_NNO}. More recently, infinite-layers nickelates were also realized by molecular beam epitaxy (MBE), using an \textit{in situ} thermal hydrogen cracker \cite{Parzyck2024_Chargeorder} or a redox-process through a metal overlayer to reduce the perovskite phase \cite{Wei2023_reduction_Al, Wei2023_SC_NEuNO}, as in the case of MBE-grown superconducting Nd$_{1-x}$Eu$_{x}$SrNiO$_{2}$ samples capped by an Al film \cite{Wei2023_SC_NEuNO}. While superconducting infinite-layer nickelates were also successfully synthesized without any capping layer \cite{Zeng2022_SC_NSNO}, superconductivity was found particularly robust in films protected by STO, which might have an important role in the reduction mechanism during the topotactic process. In particular, recent studies demonstrated that the last few surface unit-cells of uncapped NdNiO$_{2}$ (NNO) films might not be fully reduced, with excess oxygen ions arranged in a (short range) ordered pattern characterized by 1/3 reciprocal lattice units in plane periodicity \cite{Raji2023_TEM_Superstructure, Parzyck2024_Chargeorder}, possibly explaining some of the earlier evidence of charge order by resonant inelastic x-ray scattering measurements \cite{Krieger2022_Labpaper, Tam2022_CDW, Rossi2022_CO_Reciprocal}.

\par From these studies it emerges that the intrinsic electronic properties of nickelates cannot be easily accessed by surface-sensitive techniques, in particular, angle-resolved photoemission spectroscopy (ARPES) and scanning tunnelling spectroscopy (STS). It is then quite crucial to establish methods to restore the surface properties of these compounds in order to fully study their electronic properties. Among the important questions to be undisclosed, we cite the need to establish the role of rare-earth 5\textit{d}-electronic states\cite{Hepting2020_XAS_Ni2p}, the classification of the parent compounds as charge or Hubbard insulators \cite{Karp2020_Chargehubbard}, and finally a clarification of the magnetic properties that might or might not be significantly different from cuprates \cite{Fowlie2022_Magnetic_NNO}. 

\par  Here, we report a thorough study of the surface electronic properties of STO-capped and uncapped undoped NNO films realized by CaH$_{2}$ topotactic reduction of the perovskite phase. We studied films realized at the CNRS-IPCMS in Strasbourg by the PLD technique followed by the topotactic reduction, and then introduced from the ambient atmosphere into surface analysis set-ups: lab-based x-ray photoemission spectroscopy (XPS) and scanning tunnelling microscopy (STM) at the CNR-SPIN Modular Oxide Deposition and analysis (MODA) Facility, and high-resolution valence band (VB) photoemission spectroscopy at the Cassiopee beamline of the SOLEIL synchrotron radiation facility. 

\par We show that as received uncapped samples have insulating surfaces, which contain a substantial fraction of Ni in nominal Ni$^{2+}$ electronic configuration, while in capped NdNiO$_{2}$ Ni is mostly in Ni$^{1+}$ valence state. However, the unit-cells at the interface with the STO overlayer of capped NdNiO$_{2}$ are also not ideal, with some Ni in higher valence state and surprisingly Nd 3\textit{d}$_{5/2}$ core level spectra having a relevant contribution from ligand 4\textit{f}$^{4}$\underline{L} states, thus differing from ideal Nd$^{3+}$. 

\par A metallic state with a robust Ni$^{1+}$ valence in uncapped NNO is nevertheless achieved by ultra high vacuum (UHV) annealing at 250~$^{\circ}$C, while the removal of the STO-capping-layer and of the interfacial NNO-layer by unit-cell by unit-cell Ar-ion sputtering in capped NNO gives rise to a metallic surface state and a substantial recover of Nd 3\textit{d}$_{5/2}$ spectra towards ideal Nd$^{3+}$. 
The results provide insights into the properties of infinite-layer NdNiO$_{2}$ thin films prepared by the CaH$_{2}$ topotactic reduction of perovskite NdNiO$_{3}$, and suggest methods to improve their surface quality. 

\section{Experimental details}
\par Perovskite NdNiO$_3$ (NNO3) thin films of 10 nm in thickness were grown onto STO single crystal by PLD as described in Refs. \cite{Preziosi2017_Synthesis_NNO, Krieger2022_synthesis_NNO}.
In this work, the capping layer was composed of 3 STO unit cells \textit{in situ} grown epitaxially on the NNO3 film. Both capped and uncapped NNO films were then realized by topotactic reduction processes using metal hydride (CaH$_2$) as a reducing reagent \cite{Krieger2022_synthesis_NNO}.
The surface/interface valence states of uncapped and capped NNO were studied at room temperature by XPS (Scienta Omicron) using an Al K$_\alpha$ x-ray source (h$\nu$=1486.6 eV), and a hemispherical energy analyzer, operated in large area mode with a pass energy of 50 eV. The angle between the source and the analyzer was 45$^{\circ}$. In most of the measurements, the sample surface-normal was at 45$^{\circ}$ to the source and the analyzer entrance slits, to maximize the intensity. Binding energies were determined by using  Carbon-1\textit{s} core-level as reference. The XPS spectra were corrected by a Shirley background and fitted using pseudo-Voigt functions by CASA-XPS software \cite{Fairley2021_CASA}. 
Room temperature UHV-STM was carried out \textit{in situ} using a Scienta Omicron VT-AFM, using electrochemically etched tungsten and iridium tips. The samples were also studied after \textit{in situ} UHV ( pressure $<$ 5$\times$ 10$^{-10}$ Torr ) annealing  at 250~$^{\circ}$C for 12h. 
Finally, we studied the valence band of STO-capped NNO samples by photoelectron spectroscopy at 20 K at the SOLEIL Cassiopee beamline after removing the capping layer by Ar$^+$ ion sputtering. Before the introduction into the UHV photoemission system, the sample surfaces were cleaned by oxygen plasma to remove the adventitious carbon layer. In order to minimize sample damage, Ar-implantation and off-stoichiometry, we employed low energy accelerated Ar$^+$ ions of 800 eV, Ar-pressure of 10$^{-6}$ mbar and ion current of 40 mA. Each cycle time was about 10 minutes. After 6 cycles, the 3 unit cells of the STO capping layer were completely removed and the underneath pristine infinite-layer nickelate was exposed at the surface. The VB spectra were acquired using horizontal polarized 600 eV photons at an incident angle of 20$^{\circ}$, with the sample in the vertical plane, and an energy resolution of 50 meV. The Fermi energy was calibrated by using a reference gold sample.

\section{Results and discussion}

\begin{figure}
\includegraphics[width=8.7cm]{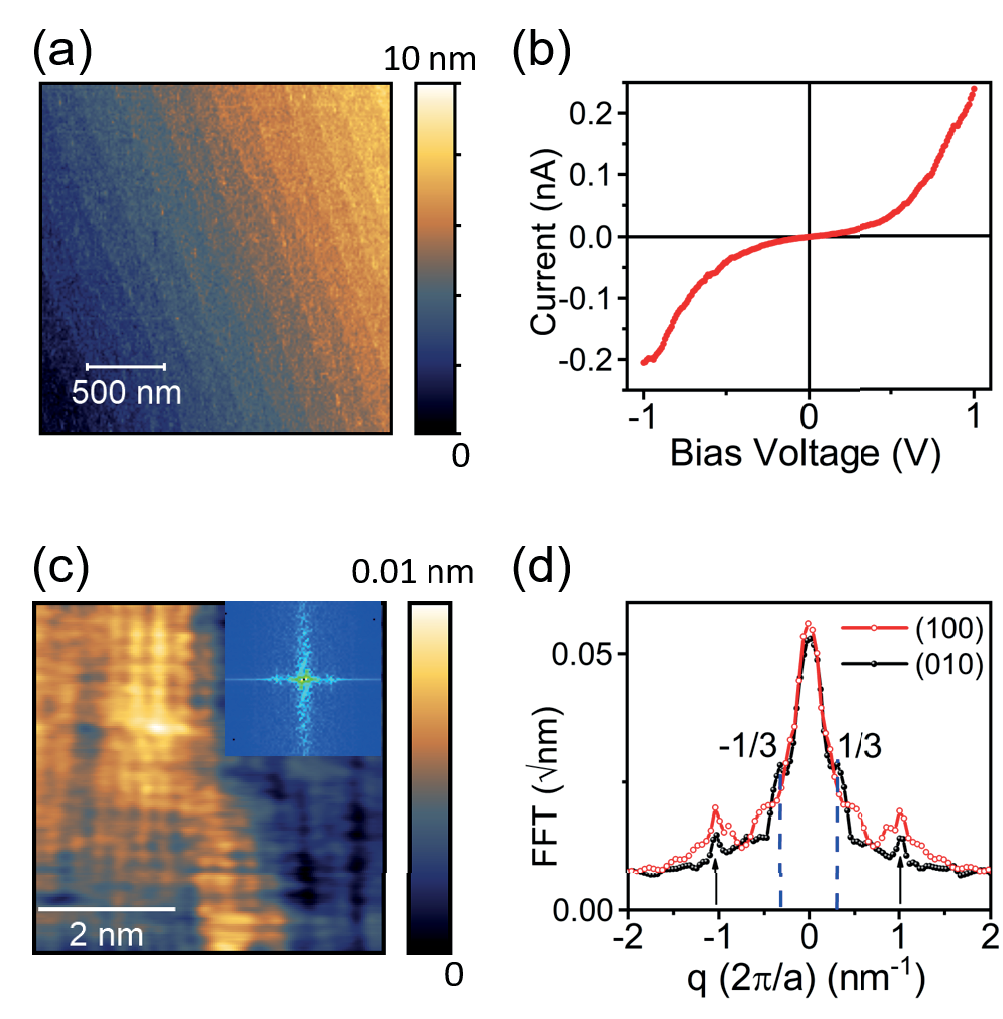}
\caption{STM measurements on uncapped nickelates: (a) a large area STM image acquired on uncapped NNO showing step-terraces (V$_{gap}$= 1.5 V I$_{tunnel}$=0.5 nA); (b) typical local STM spectroscopic I-V data of the as-grown infinite layer uncapped NNO thin film;(c) high resolution image showing the presence of atomic rows (V$_{gap}$= 1.0 V I$_{tunnel}$=0.5 nA). The inset in (c) is the Fast Fourier Transform (FFT) of the image; (d) Averaged profiles along (100) and (010) directions of the square-root FFT magnitude shown.}
\label{fig1}
\end{figure}

\begin{figure}
\includegraphics[width=8.7cm]{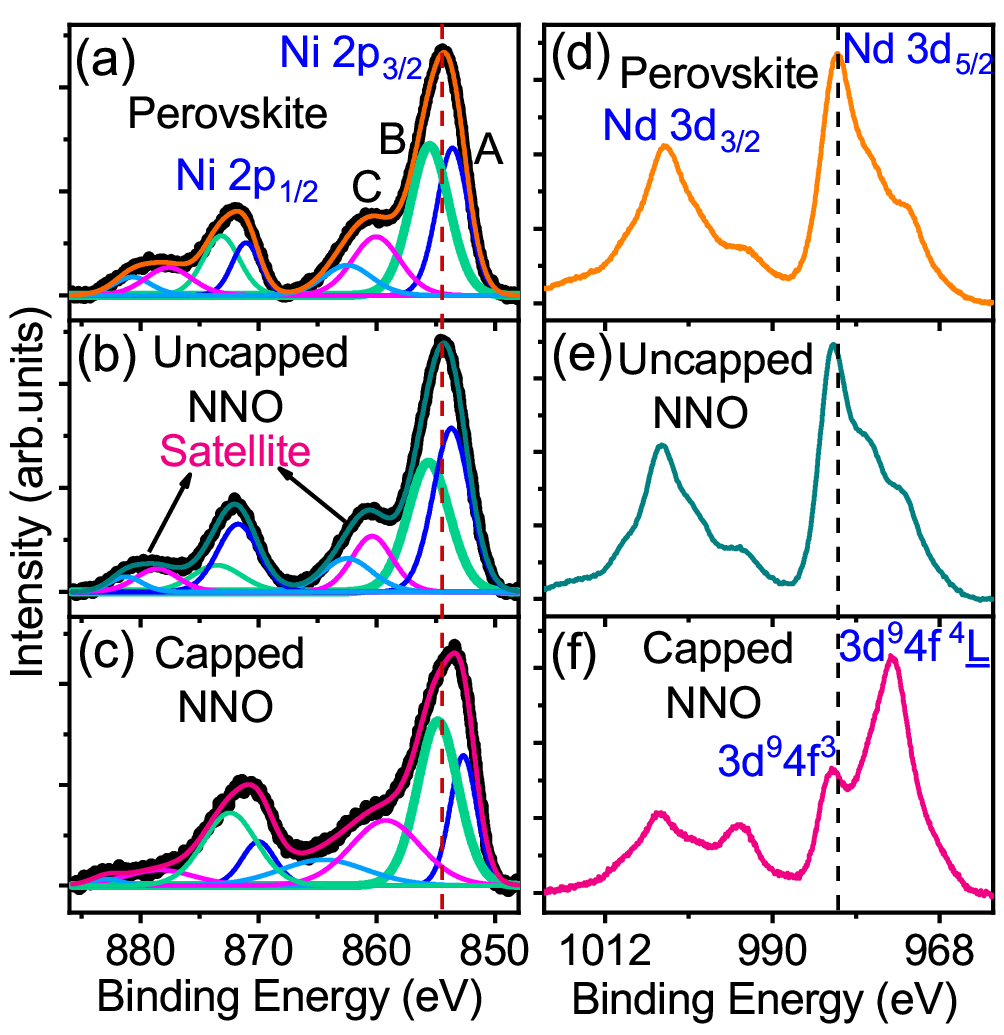}
\caption{Core-level Ni 2\textit{p} and Nd 3\textit{d} XPS spectra after subtraction of a Shirley background: (a) and (d) Perovskite; (b) and (e) uncapped NNO; (c) and (f) capped NNO. Features A, B and C in the 2\textit{p}$_{3/2}$ spectra are described in the main text. Dashed vertical lines (a)-(c) and (d)-(f) are guides for the eye showing how the main 2\textit{p}$_{3/2}$ and 3\textit{d}$^{9}$4\textit{f}$^{3}$ peaks shift going from the perovskite to the infinite-layer nickelate samples.
}
\label{fig2}
\end{figure}
 
\par To study the surface structural properties of NNO samples, we performed UHV-STM measurements on uncapped samples. In Fig. \ref{fig1}, we show a large area STM image acquired on the uncapped NNO. Without any surface treatment, stable tunnelling conditions could be achieved only by gap voltages larger than 1 V and tunnelling currents below 1 nA. This result alone suggests that the surface layer is not metallic, as confirmed by local tunnelling spectroscopy showing semiconducting I-V spectra  (Fig. \ref{fig1}b). The large area STM image in Fig. \ref{fig1}a) shows hints of step-terraces, in agreement with previous AFM data on similar samples \cite{Krieger2022_synthesis_NNO}. On the other hand, the STM data, taken in constant current mode, show also the presence of elongated islands within each terrace, \textit{i.e.} the surface is not atomically smooth. To better visualize any possible periodic pattern we show a high resolution topographic image in Fig. \ref{fig1}c, its Fast Fourier Transform (FFT) ( inset of Fig. \ref{fig1}c) and the averaged profiles along (100) and (010) directions of the square-root FFT magnitude (Fig. \ref{fig1}d) .  High resolution topography shows locally on top of most of these islands, and in-between them, a periodic pattern along the (001) direction with a spacing between consecutive rows of $0.38 \pm 0.01$ nm, while along each row (010 direction), ordered atomic lines are unclear, and atomic-resolution 1x1 unit-cells are only locally observed. The FFT shows indeed that the surface lattice does not have a 1x1 long range order, while hints of a 1/3 pattern is also observed, suggesting a link with the HRTEM reports in refs \cite{Raji2023_TEM_Superstructure, Parzyck2024_Chargeorder}. This result was found in several areas of the sample, and on different uncapped NNO realized using the same method. 

\par To get insight into the electronic properties of infinite-layers nickelates, and the effect of the topotactic process, in Figure \ref{fig2} we compare core-level Ni 2\textit{p} and Nd 3\textit{d} XPS spectra of as-grown perovskite NNO3 (Fig. \ref{fig2}a,d), uncapped (Fig. \ref{fig2}b,e) and capped (Fig. \ref{fig2}c,f) NNO samples. 
The core-level Ni 2\textit{p} spectra consist in 2\textit{p}$_{1/2}$ ($\sim$ 871 eV) and 2\textit{p}$_{3/2}$ features ($\sim$ 855 eV), composed by a lower-binding energy main peak and a higher energy satellite. The high energy satellite (labelled C) is due to final states related to fully unscreened 2\textit{p}$^5$ core-hole. The main peak, on the other hand, is related to final states where the 2\textit{p}$^5$ core-hole is screened by electrons belonging to neighbour ligand oxygen ions, or by electrons coming from neighbouring clusters \cite{Veenendaal1993_XPS_Ni2p, Veenendaal2006_Ni2p_nonlocal}. As a consequence, the 2\textit{p}$_{3/2}$ XPS spectra are decomposed in two peaks labelled A (non-local screened final states) and B (local screened final states). The fitting analysis shows that for the reference perovskite NNO3, local and non-local screening channels have similar spectral weights in good agreement with previous reports \cite{Preziosi2017_Synthesis_NNO,fu2020corelevel_XPS_NNO}. 
Since the perovskite NNO3 is regarded as a negative charge transfer system according to the Zaanen-Sawatzky-Allen scheme, at room temperature the ground state has mainly a 3\textit{d}$^{8}$\underline{L} configuration (\underline{L} denotes a ligand hole), with nominal Ni valence close to Ni$^{2+}$ \cite{Bisogni2016_RIXS_CT}. 
In infinite-layer nickelates, substantial changes in the energy and the spectral weights of A, B and C features were expected, in particular a shift of the main peak towards lower binding energy due to the lower overall oxidation state. However, in uncapped NNO, very surprisingly, the XPS Ni 2\textit{p} core-level looks very similar to the perovskite (Fig. \ref{fig2}b), with only a few differences in the A/B ratio. This result suggests that, within the probing depth of 1.5 nm, uncapped NNO are barely reduced compared to the perovskite. On the other hand, capped NNO shows a clear shift towards lower binding energy of all features and substantial changes in the relative intensities (Fig. \ref{fig2}c). Additionally, the unscreened satellite peak (C)  is characterized by a substantial decrease of spectral weight and shifts toward lower binding energies as-well, which indicates a lowering of Ni valence towards Ni$^{1+}$ \cite{Higashi2021_XPS_XAS_Ni2p,fu2020corelevel_XPS_NNO}. 
The differences in the surface and interfacial Ni-valence between uncapped and capped nickelates confirm earlier X-ray absorption spectroscopy data \cite{ Krieger2022_Labpaper}.
\par While a full understanding of the core-level Ni 2\textit{p} spectra is beyond the scope of this work requiring sophisticated theoretical calculations, due to the formal Ni 3\textit{d}$^{9}$ ground state configuration of infinite-layer nickelates, it is reasonable to look for analogies between XPS Cu core-level data of undoped cuprates and those of the capped NNO sample. The latter exhibit a valence closer to Ni$^{1+}$, thus a Ni 3\textit{d}$^{9}$ configuration similar to Cu$^{2+}$ in the CuO$_{2}$ planes. 
In cuprates, the features B and A in the 2\textit{p}$_{3/2}$ core level XPS have been interpreted as 2\textit{p}$^5$3\textit{d}$^{10}$\underline{L} local screened, and 2\textit{p}$^5$3\textit{d}$^{10}$\underline{Z} non-local screened final states, where \underline{Z} stands as Zhang-Rice Singlet (ZRS) \cite{Taguchi2005_XPS_Cu_ZRS}. 
In the photo-excitation process, ZRS is an energetically favoured bound state between non-local electron screened hole and the pristine hole in neighbour clusters. 
Multiple cluster calculations are able to catch some general characteristics of the cuprate core-level spectra and in particular the relative intensities of features A and B \cite{Veenendaal1994_XPS_Cu_Theory}. It turns out that in cuprates the low-energy feature A dominates over feature B as a consequence of the low charge transfer energy, which favours the formation of ZRS bound states. On the other hand, since the charge transfer energy in infinite-layer nickelates is inherently larger, it is unclear if ZRS bound states are effectively stable or energetically favourable. As a matter of fact, the ratio between A and B spectral weights shown in Fig. \ref{fig2}c (bottom panel) in capped NNO is opposite to that of undoped cuprates, \textit{i.e.} the low binding energy feature A has lower spectral weight compared to feature B. This result was also observed at the bare surface of NNO single crystals in refs. \cite{fu2020corelevel_XPS_NNO, Higashi2021_XPS_XAS_Ni2p}
and interpreted as a further indication that NNO is indeed a Hubbard and not a charge transfer insulator.

\par In order to get more insight into the electronic properties of as-grown capped and uncapped NNO, we performed core-level XPS Nd 3\textit{d} measurements and compared the results to perovskite NNO3 samples, as shown in Figures \ref{fig2}d-f. The main core-level peaks at $\sim$ 982 and $\sim$ 1004.2 eV correspond to Nd 3\textit{d}$_{5/2}$ and 3\textit{d}$_{3/2}$ states, respectively, and reflect the Nd$^{3+}$ oxidation state. Perovskite and uncapped NNO3 are characterized by similar Nd 3\textit{d} core-level features, slightly shifted to higher binding energy by 0.9 eV in the case of uncapped NNO in good agreement with previous report on NNO single crystals \cite{fu2020corelevel_XPS_NNO}. On the other hand, capped NNO shows a very different Nd 3\textit{d}$_{5/2}$ feature, with 3\textit{d}$^{9}$4\textit{f}$^{3}$ final state reduced in comparison to uncapped NNO, and a strongly enhanced 3\textit{d}$^{9}$4\textit{f}$^{4}$\underline{L} final state. This result shows that few NNO unit cells (1 or 2) at the interface with the STO capping layer are not ideal. There could be several reasons why the upper NiO$_2$/TiO$_2$ interface cannot be intrinsically sharp and ideal. For example, it is known that a NdTiNiO$_{3}$ intermixed phase is created at the interface between NNO and the STO-single crystal substrate due to the atomic reconstruction that compensates the interfacial polar discontinuity, as demonstrated by combined high-resolution transmission electron microscopy (HR-TEM) and ab-initio calculations \cite{Goodge2023_NNO_Interface_TEM}. As a consequence, it is not inconceivable to expect that a similar intermixed phase, maybe just including the interface, is created at the top NNO/STO interface. Furthermore, we do not exclude a possible accumulation of hydrogen at the interface after the topotactic process, which might help in explaining the anomalous Nd$^{3+}$ core-level spectrum in as-received capped reduced nickelate films.   

\begin{figure}
\includegraphics[width=8.7 cm]{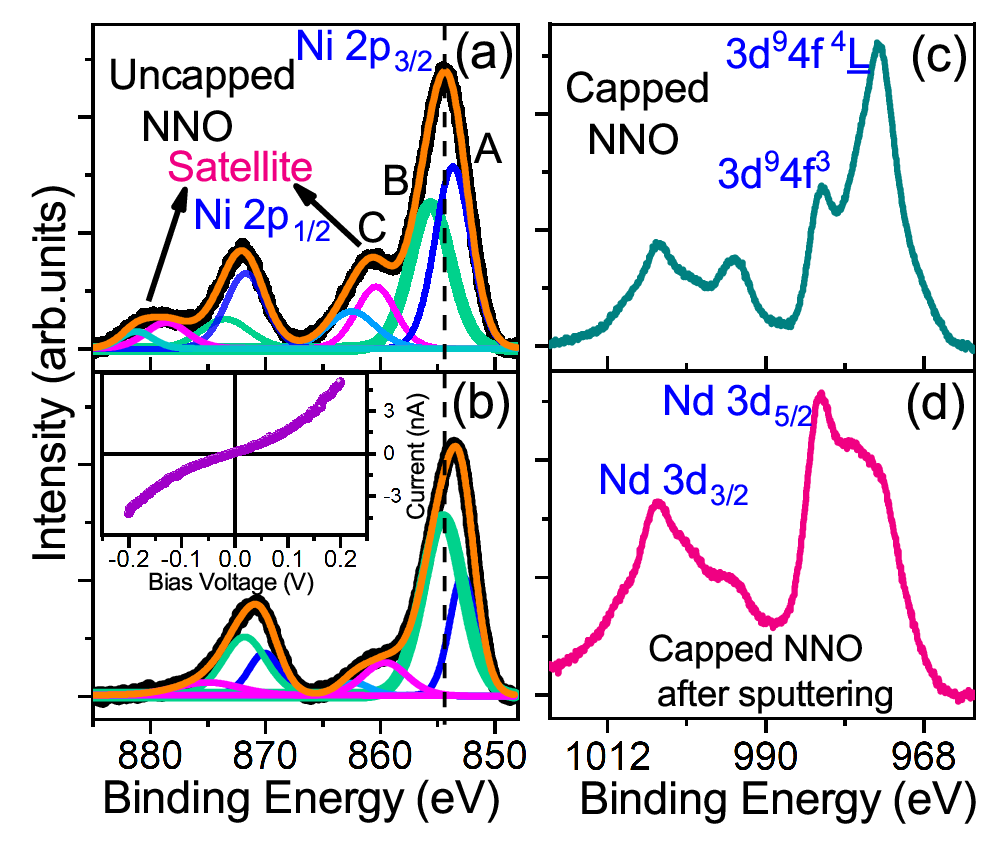}
\caption{Ni 2\textit{p} XPS core-level spectra of (a) as-grown and (b) UHV annealed uncapped NNO. The black dashed line indicates the position of the core-level Ni 2\textit{p} peak shifted to the lower binding energy after vacuum annealing. Inset in Fig. \ref{fig4}(b) shows STM local I-V spectroscopy of the uncapped NNO after annealing (V$_{gap}$=0.2 V I$_{tunnel}$=1 nA). Further high resolution XPS main-line Nd 3\textit{d} spectra of (c) As-grown NNO samples with STO capping layer and (d) After etching the capping layer by Ar$^+$ sputtering with low energy of 800 eV with different cycles for 1h inside the analyzer chamber. The black dashed line indicates the enhancement of 3\textit{d}$^{9}$4\textit{f}$^{3}$ compared to 3\textit{d}$^{9}$4\textit{f}$^{4}$\underline{L} peak in the sputtered nickelates sample.}
\label{fig3}
\end{figure}

\par From the STM and XPS studies of untreated (uncapped and capped) NNO samples it clearly emerges that their surface and interface unit cells do not have an ideal NNO structure and stoichiometry, thus any surface/interface probe of their electronic properties would not be able to provide accurate insights about the differences between cuprates and nickelates. In the following, we suggest possible directions to address this problem. 
\par First, we studied the non-ideal and barely reduced surface of uncapped NNO. To reduce excess oxygen and the amount of adsorbed species due to the \textit{ex-situ} exposure to the atmosphere from the surface unit cells of uncapped NNO, the films were annealed \textit{in situ} in UHV at 250 $^{\circ}$C for 12h, as described above. Figure \ref{fig3} shows XPS Ni 2\textit{p} core-level spectra (Fig. \ref{fig3}a before and Fig. \ref{fig3}b after annealing). We do see pronounced changes in the core-level spectra: first, the main core-level features in the annealed sample shift by 0.9 eV towards lower binding energy; second, the spectral weight of the non-local screened feature A becomes (much) lower than the local screened final state B; finally, the intensity of the unscreened satellite peak (C) is strongly suppressed after annealing \cite{fu2020corelevel_XPS_NNO, Higashi2021_XPS_XAS_Ni2p}. 
The results suggest the reduction of the Ni$^{2+}$ valence of as-grown uncapped NNO surface towards Ni$^{1+}$ in UHV annealed NNO. 
At the same time, we find that the surface is characterized by a metallic surface state, as shown by \textit{in situ} STM local I-V spectroscopy (inset in Fig. \ref{fig3}b). It is worth noting that local superconducting gap features in uncapped Sr-doped NNO were measured in uncapped samples after UHV annealing at a slightly reduced temperature in ref. \cite{Gu2020_STM_NNSO}. However, annealing in UHV, while capable of recovering a metallic surface state and a Ni$^{1+}$ oxidation state, is unable to change the overall morphology and to create a long-range ordered surface lattice. The irregular morphology on top of the terraces could not be modified substantially by the low temperature UHV annealing process.
\par We now go back to the capped NNO samples. To investigate the NNO electronic properties in capped NNO samples, and to possibly remove the non-ideal interfacial layer, we used gentle Ar$^+$ ion-beam etching at the Cassiopee beamline, as described in the experimental details section. In Figures \ref{fig3}c,d we show the effect of Ar$^+$ ion etching removal of the STO capping-layer on the Nd 3\textit{d} spectra, comparing XPS data acquired before (\ref{fig3}c) and after (\ref{fig3}d) the capping removal. We can see that after etching, the NNO-surface state shows a Nd$^{3+}$ spectra characterized by a substantial reduction of the prominent 3\textit{d}$^{9}$4\textit{f}$^{4}$\underline{L} final state feature observed before Ar$^+$ ion etching. This result confirms that some kind of intermixed phase was formed at the NNO/STO interface. 
At the same time the Ni 2\textit{p} spectrum (not shown) is essentially not affected by the Ar-ion etching process, \textit{i.e.} the surface of the NNO sample after capping removal keeps the overall Ni$^{1+}$ valence state. 

\begin{figure}
\includegraphics[width=8.7 cm]{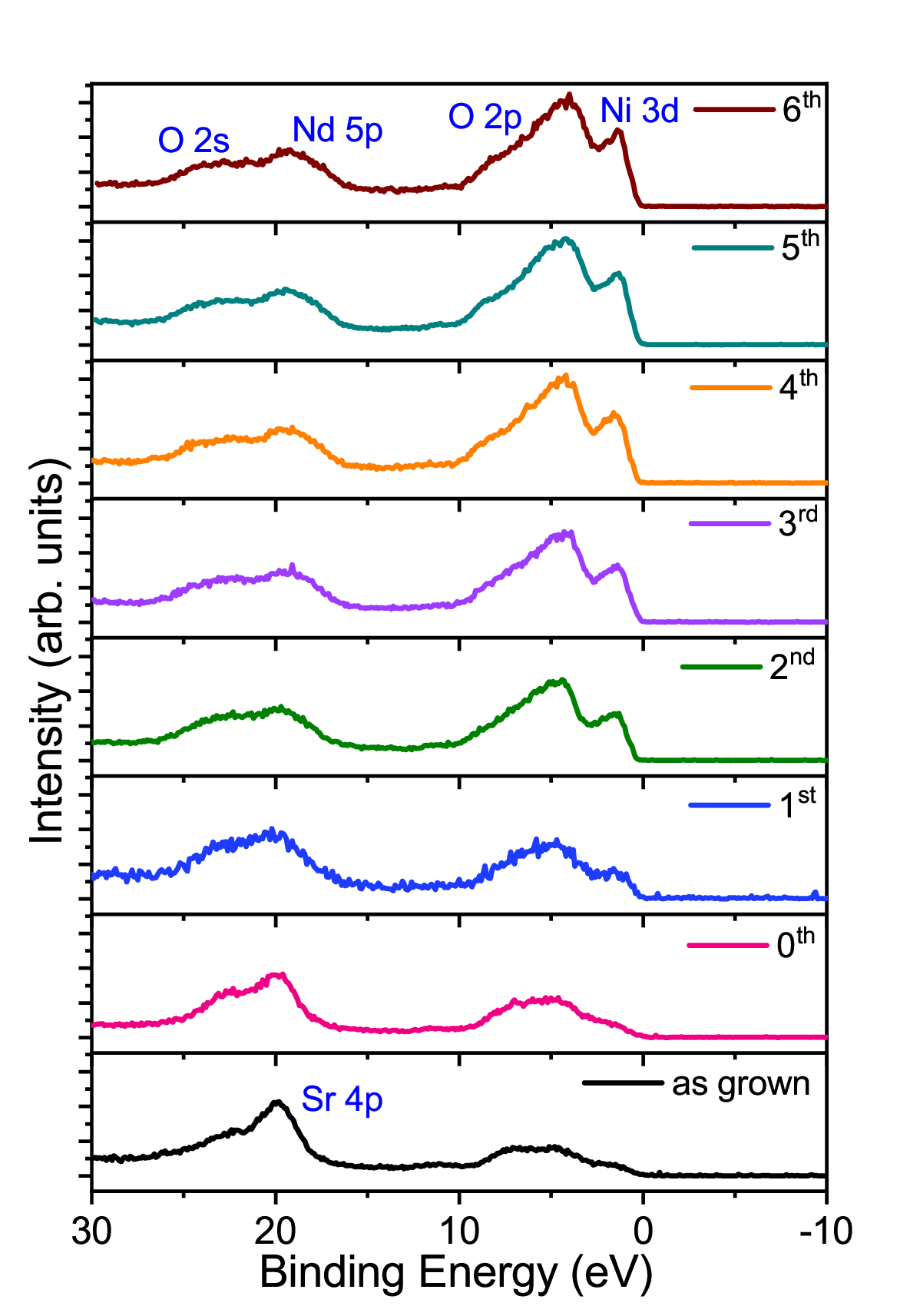}
\caption{High resolution Valence Band spectra of capped-NNO after different cycles of Ar-ion etching obtained at the Cassiopee beamline. The spectra show, together with the VB region, core-level Nd 5\textit{p}, Sr 4\textit{p}. The VB spectra (E$_{F}$ to -10 eV) are composed of O 2\textit{p}, Ni 3\textit{d} and Ti 3\textit{d} contribution.}
\label{fig4}
\end{figure}

\par Figure \ref{fig4} shows the valence band (VB) spectra of the capped NNO sample acquired with an incident photon energy of 600 eV and a resolution of 50 meV, including the VB (E$_F$ to -10 eV), core level Nd 5\textit{p} and Sr 4\textit{p} features after different cycles of Ar$^+$ sputtering. In the pristine state, the spectrum shows mostly the O 2\textit{p} VB of the topmost STO, with possible traces of in-gap Ti 3\textit{d} and Ni 3\textit{d} features below E$_{F}$. Moreover, the Sr 4\textit{p} peak is well visible around 20 eV of binding energy. After a few cycles of Ar$^+$ ion etching, and at the end of the 6$^{th}$ cycle of sputtering, the Sr 4\textit{p} (as well as all core levels related to Ti and Sr ions, not shown)  is gradually suppressed, a Nd 5\textit{p} feature around 18 eV binding energy appears, and the VB changes substantially. At the end of 6$^{th}$ cycle, the STO capping is removed together with the first intermixed NNO-STO unit cell. At the same time, the O 2\textit{p} VB changes completely in shape, with a peak located between 6-8 eV, and Ni 3\textit{d} states crossing the Fermi level show up.
The final VB spectra after Ar$^+$ sputtering have similarities with that one of uncapped PrNiO$_{2}$ measured by soft X-ray photoemission spectroscopy, showing the appearance of Ni 3\textit{d} states crossing the Fermi-level \cite{CHEN20221806_XPS_VB_Ni3d}. On the other hand, the VB peak is more intense than the Ni 3\textit{d} feature. This can be possibly related to the contribution to the VB of Nd 5\textit{d} and Nd 4\textit{f} states together with O 2\textit{p} states. From the data, we can tentatively estimate the differences in binding energy between O 2\textit{p} and Ni 3\textit{d} electronic states (E$_{O 2\textit{p}}$-E$_{Ni 3\textit{d}}$), which is found to be 2.72 eV, matching the values reported for PrNiO$_{2}$ in ref.\cite{CHEN20221806_XPS_VB_Ni3d}. However, the contribution to the VB from states other than O 2\textit{p} makes this estimation somewhat uncertain. Assuming the value of (E$_{O 2\textit{p}}$-E$_{Ni 3\textit{d}}$) found correct, following ref. \cite{CHEN20221806_XPS_VB_Ni3d} the result suggests a Hubbard energy U lower than the charge transfer energy $\Delta$ (U$ \leq$ $\Delta$).

\section{Discussion and conclusions}
\par In this work, we investigated the surface/interface electronic properties of uncapped and capped infinite-layer nickelates. The experimental results of our combined STM and photoemission spectroscopy studies show that the surface and interface unit cells of uncapped and capped NNO are not ideal. While uncapped NNO is characterized by an insulating surface state and a Ni-valence remarkably similar to reference NNO3 perovskite, capped NNO shows mostly Ni in Ni$^{1+}$ configuration. On the other hand, at the interface with the STO capping an intermixed non-ideal layer, presumably of NdTiNiO$_{x}$ composition (and possibly excess hydrogen), is formed. A metallic surface with a Ni$^{1+}$ valence state was restored via annealing in UHV at 250 $^{\circ}$C in uncapped NNO, while an Ar$^+$ ion sputtering process was necessary to access the electronic properties of capped NNO samples by removing the STO layers and the interfacial (non-ideal) unit cell. In this way, we could provide an experimental study of the VB of Ni$^{1+}$ NdNiO$_{2}$ nickelate and access the near Fermi level electronic properties, giving further support to the classification of these oxides as Hubbard or intermediate between Hubbard and charge transfer insulators. 

Our work represents a step towards the goal of creating an ideal NdNiO$_2$ surface state, which could allow deep investigations by surface spectroscopy methods like angle-resolved photoemission. While annealing in UHV and ion-sputtering are not yet sufficient to establish long-range atomically flat ideal NiO$_2$ surfaces on samples prepared by ex-situ CaH$_{2}$ topotactic reduction of perovskite NdNiO$_{3}$ samples, their opportune combination and optimization are promising for future studies aimed at preserving not only the stoichiometry but also the long-range surface structural order. These methods combined with the novel \textit{in-situ} approaches of realization of infinite layer nickelates, like hydrogen thermal cracker technique \cite{Parzyck2024_Chargeorder} and metal overlayer deposition \cite{ Wei2023_reduction_Al}, or the investigation of other epitaxially-grown capping-layers seem promising in the direction to improve the surface quality of infinite-layer nickelates.  

\begin{acknowledgments}
This work was supported by the European Union's Horizon 2022 research and innovation programme under the Marie Sklodowska-Curie project MODERN, grant agreement No. 101108695, and by the Ministry of University and Research project PRIN QTFLUO No.20207ZXT4Z. This work was funded by the French National Research Agency (ANR) through the ANR-JCJC FOXIES ANR-21-CE08-0021. This work was also done as part of the Interdisciplinary Thematic Institute QMat, ITI 2021 2028 program of the University of Strasbourg, CNRS and Inserm, and supported by IdEx Unistra (ANR 10 IDEX 0002), and by SFRI STRAT’US project (ANR 20 SFRI 0012) and EUR QMAT ANR-17-EURE-0024 under the framework of the French Investments for the Future Program. The synchrotron experiments were performed at SOLEIL under proposal number 20220304, and the staff of Cassiopee beamline is acknowledged for their technical support. We acknowledge Prof. G. Ghiringhelli for the discussions on the physics of nickelates. 
\end{acknowledgments}


\begin{thebibliography}{26}%
\makeatletter
\providecommand \@ifxundefined [1]{%
 \@ifx{#1\undefined}
}%
\providecommand \@ifnum [1]{%
 \ifnum #1\expandafter \@firstoftwo
 \else \expandafter \@secondoftwo
 \fi
}%
\providecommand \@ifx [1]{%
 \ifx #1\expandafter \@firstoftwo
 \else \expandafter \@secondoftwo
 \fi
}%
\providecommand \natexlab [1]{#1}%
\providecommand \enquote  [1]{``#1''}%
\providecommand \bibnamefont  [1]{#1}%
\providecommand \bibfnamefont [1]{#1}%
\providecommand \citenamefont [1]{#1}%
\providecommand \href@noop [0]{\@secondoftwo}%
\providecommand \href [0]{\begingroup \@sanitize@url \@href}%
\providecommand \@href[1]{\@@startlink{#1}\@@href}%
\providecommand \@@href[1]{\endgroup#1\@@endlink}%
\providecommand \@sanitize@url [0]{\catcode `\\12\catcode `\$12\catcode
  `\&12\catcode `\#12\catcode `\^12\catcode `\_12\catcode `\%12\relax}%
\providecommand \@@startlink[1]{}%
\providecommand \@@endlink[0]{}%
\providecommand \url  [0]{\begingroup\@sanitize@url \@url }%
\providecommand \@url [1]{\endgroup\@href {#1}{\urlprefix }}%
\providecommand \urlprefix  [0]{URL }%
\providecommand \Eprint [0]{\href }%
\providecommand \doibase [0]{http://dx.doi.org/}%
\providecommand \selectlanguage [0]{\@gobble}%
\providecommand \bibinfo  [0]{\@secondoftwo}%
\providecommand \bibfield  [0]{\@secondoftwo}%
\providecommand \translation [1]{[#1]}%
\providecommand \BibitemOpen [0]{}%
\providecommand \bibitemStop [0]{}%
\providecommand \bibitemNoStop [0]{.\EOS\space}%
\providecommand \EOS [0]{\spacefactor3000\relax}%
\providecommand \BibitemShut  [1]{\csname bibitem#1\endcsname}%
\let\auto@bib@innerbib\@empty
\bibitem [{\citenamefont {Anisimov}\ \emph {et~al.}(1999)\citenamefont
  {Anisimov}, \citenamefont {Bukhvalov},\ and\ \citenamefont
  {Rice}}]{Anisimov1999_Per_nickelates}%
  \BibitemOpen
  \bibfield  {author} {\bibinfo {author} {\bibfnamefont {V.~I.}\ \bibnamefont
  {Anisimov}}, \bibinfo {author} {\bibfnamefont {D.}~\bibnamefont {Bukhvalov}},
  \ and\ \bibinfo {author} {\bibfnamefont {T.~M.}\ \bibnamefont {Rice}},\
  }\href {\doibase 10.1103/PhysRevB.59.7901} {\bibfield  {journal} {\bibinfo
  {journal} {Phys. Rev. B}\ }\textbf {\bibinfo {volume} {59}},\ \bibinfo
  {pages} {7901} (\bibinfo {year} {1999})}\BibitemShut {NoStop}%
\bibitem [{\citenamefont {Li}\ \emph {et~al.}(2019)\citenamefont {Li},
  \citenamefont {Lee}, \citenamefont {Wang}, \citenamefont {Osada},
  \citenamefont {Crossley}, \citenamefont {Lee}, \citenamefont {Cui},
  \citenamefont {Hikita},\ and\ \citenamefont {Hwang}}]{Li2019_SC_NNO}%
  \BibitemOpen
  \bibfield  {author} {\bibinfo {author} {\bibfnamefont {D.}~\bibnamefont
  {Li}}, \bibinfo {author} {\bibfnamefont {K.}~\bibnamefont {Lee}}, \bibinfo
  {author} {\bibfnamefont {B.~Y.}\ \bibnamefont {Wang}}, \bibinfo {author}
  {\bibfnamefont {M.}~\bibnamefont {Osada}}, \bibinfo {author} {\bibfnamefont
  {S.}~\bibnamefont {Crossley}}, \bibinfo {author} {\bibfnamefont {H.~R.}\
  \bibnamefont {Lee}}, \bibinfo {author} {\bibfnamefont {Y.}~\bibnamefont
  {Cui}}, \bibinfo {author} {\bibfnamefont {Y.}~\bibnamefont {Hikita}}, \ and\
  \bibinfo {author} {\bibfnamefont {H.~Y.}\ \bibnamefont {Hwang}},\ }\href
  {\doibase 10.1038/s41586-019-1496-5} {\bibfield  {journal} {\bibinfo
  {journal} {Nature}\ }\textbf {\bibinfo {volume} {572}},\ \bibinfo {pages}
  {624} (\bibinfo {year} {2019})}\BibitemShut {NoStop}%
\bibitem [{\citenamefont {Parzyck}\ \emph {et~al.}(2024)\citenamefont
  {Parzyck}, \citenamefont {Gupta}, \citenamefont {Wu}, \citenamefont {Anil},
  \citenamefont {Bhatt}, \citenamefont {Bouliane}, \citenamefont {Gong},
  \citenamefont {Gregory}, \citenamefont {Luo}, \citenamefont {Sutarto},
  \citenamefont {He}, \citenamefont {Chuang}, \citenamefont {Zhou},
  \citenamefont {Herranz}, \citenamefont {Kourkoutis}, \citenamefont {Singer},
  \citenamefont {Schlom}, \citenamefont {Hawthorn},\ and\ \citenamefont
  {Shen}}]{Parzyck2024_Chargeorder}%
  \BibitemOpen
  \bibfield  {author} {\bibinfo {author} {\bibfnamefont {C.~T.}\ \bibnamefont
  {Parzyck}}, \bibinfo {author} {\bibfnamefont {N.~K.}\ \bibnamefont {Gupta}},
  \bibinfo {author} {\bibfnamefont {Y.}~\bibnamefont {Wu}}, \bibinfo {author}
  {\bibfnamefont {V.}~\bibnamefont {Anil}}, \bibinfo {author} {\bibfnamefont
  {L.}~\bibnamefont {Bhatt}}, \bibinfo {author} {\bibfnamefont
  {M.}~\bibnamefont {Bouliane}}, \bibinfo {author} {\bibfnamefont
  {R.}~\bibnamefont {Gong}}, \bibinfo {author} {\bibfnamefont {B.~Z.}\
  \bibnamefont {Gregory}}, \bibinfo {author} {\bibfnamefont {A.}~\bibnamefont
  {Luo}}, \bibinfo {author} {\bibfnamefont {R.}~\bibnamefont {Sutarto}},
  \bibinfo {author} {\bibfnamefont {F.}~\bibnamefont {He}}, \bibinfo {author}
  {\bibfnamefont {Y.-D.}\ \bibnamefont {Chuang}}, \bibinfo {author}
  {\bibfnamefont {T.}~\bibnamefont {Zhou}}, \bibinfo {author} {\bibfnamefont
  {G.}~\bibnamefont {Herranz}}, \bibinfo {author} {\bibfnamefont {L.~F.}\
  \bibnamefont {Kourkoutis}}, \bibinfo {author} {\bibfnamefont
  {A.}~\bibnamefont {Singer}}, \bibinfo {author} {\bibfnamefont {D.~G.}\
  \bibnamefont {Schlom}}, \bibinfo {author} {\bibfnamefont {D.~G.}\
  \bibnamefont {Hawthorn}}, \ and\ \bibinfo {author} {\bibfnamefont {K.~M.}\
  \bibnamefont {Shen}},\ }\href {\doibase 10.1038/s41563-024-01797-0}
  {\bibfield  {journal} {\bibinfo  {journal} {Nat. Mater.}\ } (\bibinfo {year}
  {2024}),\ 10.1038/s41563-024-01797-0}\BibitemShut {NoStop}%
\bibitem [{\citenamefont {Wei}\ \emph {et~al.}(2023{\natexlab{a}})\citenamefont
  {Wei}, \citenamefont {Shin}, \citenamefont {Hong}, \citenamefont {Shin},
  \citenamefont {Thind}, \citenamefont {Yang}, \citenamefont {Klie},
  \citenamefont {Walker},\ and\ \citenamefont {Ahn}}]{Wei2023_reduction_Al}%
  \BibitemOpen
  \bibfield  {author} {\bibinfo {author} {\bibfnamefont {W.}~\bibnamefont
  {Wei}}, \bibinfo {author} {\bibfnamefont {K.}~\bibnamefont {Shin}}, \bibinfo
  {author} {\bibfnamefont {H.}~\bibnamefont {Hong}}, \bibinfo {author}
  {\bibfnamefont {Y.}~\bibnamefont {Shin}}, \bibinfo {author} {\bibfnamefont
  {A.~S.}\ \bibnamefont {Thind}}, \bibinfo {author} {\bibfnamefont
  {Y.}~\bibnamefont {Yang}}, \bibinfo {author} {\bibfnamefont {R.~F.}\
  \bibnamefont {Klie}}, \bibinfo {author} {\bibfnamefont {F.~J.}\ \bibnamefont
  {Walker}}, \ and\ \bibinfo {author} {\bibfnamefont {C.~H.}\ \bibnamefont
  {Ahn}},\ }\href {\doibase 10.1103/PhysRevMaterials.7.013802} {\bibfield
  {journal} {\bibinfo  {journal} {Phys. Rev. Mater.}\ }\textbf {\bibinfo
  {volume} {7}},\ \bibinfo {pages} {013802} (\bibinfo {year}
  {2023}{\natexlab{a}})}\BibitemShut {NoStop}%
\bibitem [{\citenamefont {Wei}\ \emph {et~al.}(2023{\natexlab{b}})\citenamefont
  {Wei}, \citenamefont {Vu}, \citenamefont {Zhang}, \citenamefont {Walker},\
  and\ \citenamefont {Ahn}}]{Wei2023_SC_NEuNO}%
  \BibitemOpen
  \bibfield  {author} {\bibinfo {author} {\bibfnamefont {W.}~\bibnamefont
  {Wei}}, \bibinfo {author} {\bibfnamefont {D.}~\bibnamefont {Vu}}, \bibinfo
  {author} {\bibfnamefont {Z.}~\bibnamefont {Zhang}}, \bibinfo {author}
  {\bibfnamefont {F.~J.}\ \bibnamefont {Walker}}, \ and\ \bibinfo {author}
  {\bibfnamefont {C.~H.}\ \bibnamefont {Ahn}},\ }\href {\doibase
  10.1126/sciadv.adh3327} {\bibfield  {journal} {\bibinfo  {journal} {Sci.
  Adv.}\ }\textbf {\bibinfo {volume} {9}},\ \bibinfo {pages} {eadh3327}
  (\bibinfo {year} {2023}{\natexlab{b}})}\BibitemShut {NoStop}%
\bibitem [{\citenamefont {Zeng}\ \emph {et~al.}(2022)\citenamefont {Zeng},
  \citenamefont {Yin}, \citenamefont {Li}, \citenamefont {Chow}, \citenamefont
  {Tang}, \citenamefont {Han}, \citenamefont {Huang}, \citenamefont {Cao},
  \citenamefont {Wan}, \citenamefont {Zhang}, \citenamefont {Lim},
  \citenamefont {Diao}, \citenamefont {Yang}, \citenamefont {Wee},
  \citenamefont {Pennycook},\ and\ \citenamefont {Ariando}}]{Zeng2022_SC_NSNO}%
  \BibitemOpen
  \bibfield  {author} {\bibinfo {author} {\bibfnamefont {S.~W.}\ \bibnamefont
  {Zeng}}, \bibinfo {author} {\bibfnamefont {X.~M.}\ \bibnamefont {Yin}},
  \bibinfo {author} {\bibfnamefont {C.~J.}\ \bibnamefont {Li}}, \bibinfo
  {author} {\bibfnamefont {L.~E.}\ \bibnamefont {Chow}}, \bibinfo {author}
  {\bibfnamefont {C.~S.}\ \bibnamefont {Tang}}, \bibinfo {author}
  {\bibfnamefont {K.}~\bibnamefont {Han}}, \bibinfo {author} {\bibfnamefont
  {Z.}~\bibnamefont {Huang}}, \bibinfo {author} {\bibfnamefont
  {Y.}~\bibnamefont {Cao}}, \bibinfo {author} {\bibfnamefont {D.~Y.}\
  \bibnamefont {Wan}}, \bibinfo {author} {\bibfnamefont {Z.~T.}\ \bibnamefont
  {Zhang}}, \bibinfo {author} {\bibfnamefont {Z.~S.}\ \bibnamefont {Lim}},
  \bibinfo {author} {\bibfnamefont {C.~Z.}\ \bibnamefont {Diao}}, \bibinfo
  {author} {\bibfnamefont {P.}~\bibnamefont {Yang}}, \bibinfo {author}
  {\bibfnamefont {A.~T.~S.}\ \bibnamefont {Wee}}, \bibinfo {author}
  {\bibfnamefont {S.~J.}\ \bibnamefont {Pennycook}}, \ and\ \bibinfo {author}
  {\bibfnamefont {A.}~\bibnamefont {Ariando}},\ }\href {\doibase
  10.1038/s41467-022-28390-w} {\bibfield  {journal} {\bibinfo  {journal} {Nat.
  Commun.}\ }\textbf {\bibinfo {volume} {13}},\ \bibinfo {pages} {743}
  (\bibinfo {year} {2022})}\BibitemShut {NoStop}%
\bibitem [{\citenamefont {Raji}\ \emph {et~al.}(2023)\citenamefont {Raji},
  \citenamefont {Krieger}, \citenamefont {Viart}, \citenamefont {Preziosi},
  \citenamefont {Rueff},\ and\ \citenamefont
  {Gloter}}]{Raji2023_TEM_Superstructure}%
  \BibitemOpen
  \bibfield  {author} {\bibinfo {author} {\bibfnamefont {A.}~\bibnamefont
  {Raji}}, \bibinfo {author} {\bibfnamefont {G.}~\bibnamefont {Krieger}},
  \bibinfo {author} {\bibfnamefont {N.}~\bibnamefont {Viart}}, \bibinfo
  {author} {\bibfnamefont {D.}~\bibnamefont {Preziosi}}, \bibinfo {author}
  {\bibfnamefont {J.-P.}\ \bibnamefont {Rueff}}, \ and\ \bibinfo {author}
  {\bibfnamefont {A.}~\bibnamefont {Gloter}},\ }\href {\doibase
  https://doi.org/10.1002/smll.202304872} {\bibfield  {journal} {\bibinfo
  {journal} {Small}\ }\textbf {\bibinfo {volume} {19}},\ \bibinfo {pages}
  {2304872} (\bibinfo {year} {2023})}\BibitemShut {NoStop}%
\bibitem [{\citenamefont {Krieger}\ \emph
  {et~al.}(2022{\natexlab{a}})\citenamefont {Krieger}, \citenamefont
  {Martinelli}, \citenamefont {Zeng}, \citenamefont {Chow}, \citenamefont
  {Kummer}, \citenamefont {Arpaia}, \citenamefont {Moretti~Sala}, \citenamefont
  {Brookes}, \citenamefont {Ariando}, \citenamefont {Viart}, \citenamefont
  {Salluzzo}, \citenamefont {Ghiringhelli},\ and\ \citenamefont
  {Preziosi}}]{Krieger2022_Labpaper}%
  \BibitemOpen
  \bibfield  {author} {\bibinfo {author} {\bibfnamefont {G.}~\bibnamefont
  {Krieger}}, \bibinfo {author} {\bibfnamefont {L.}~\bibnamefont {Martinelli}},
  \bibinfo {author} {\bibfnamefont {S.}~\bibnamefont {Zeng}}, \bibinfo {author}
  {\bibfnamefont {L.~E.}\ \bibnamefont {Chow}}, \bibinfo {author}
  {\bibfnamefont {K.}~\bibnamefont {Kummer}}, \bibinfo {author} {\bibfnamefont
  {R.}~\bibnamefont {Arpaia}}, \bibinfo {author} {\bibfnamefont
  {M.}~\bibnamefont {Moretti~Sala}}, \bibinfo {author} {\bibfnamefont {N.~B.}\
  \bibnamefont {Brookes}}, \bibinfo {author} {\bibfnamefont {A.}~\bibnamefont
  {Ariando}}, \bibinfo {author} {\bibfnamefont {N.}~\bibnamefont {Viart}},
  \bibinfo {author} {\bibfnamefont {M.}~\bibnamefont {Salluzzo}}, \bibinfo
  {author} {\bibfnamefont {G.}~\bibnamefont {Ghiringhelli}}, \ and\ \bibinfo
  {author} {\bibfnamefont {D.}~\bibnamefont {Preziosi}},\ }\href {\doibase
  10.1103/PhysRevLett.129.027002} {\bibfield  {journal} {\bibinfo  {journal}
  {Phys. Rev. Lett.}\ }\textbf {\bibinfo {volume} {129}},\ \bibinfo {pages}
  {027002} (\bibinfo {year} {2022}{\natexlab{a}})}\BibitemShut {NoStop}%
\bibitem [{\citenamefont {Tam}\ \emph {et~al.}(2022)\citenamefont {Tam},
  \citenamefont {Choi}, \citenamefont {Ding}, \citenamefont {Agrestini},
  \citenamefont {Nag}, \citenamefont {Wu}, \citenamefont {Huang}, \citenamefont
  {Luo}, \citenamefont {Gao}, \citenamefont {García-Fernández}, \citenamefont
  {Qiao},\ and\ \citenamefont {Zhou}}]{Tam2022_CDW}%
  \BibitemOpen
  \bibfield  {author} {\bibinfo {author} {\bibfnamefont {C.~C.}\ \bibnamefont
  {Tam}}, \bibinfo {author} {\bibfnamefont {J.}~\bibnamefont {Choi}}, \bibinfo
  {author} {\bibfnamefont {X.}~\bibnamefont {Ding}}, \bibinfo {author}
  {\bibfnamefont {S.}~\bibnamefont {Agrestini}}, \bibinfo {author}
  {\bibfnamefont {A.}~\bibnamefont {Nag}}, \bibinfo {author} {\bibfnamefont
  {M.}~\bibnamefont {Wu}}, \bibinfo {author} {\bibfnamefont {B.}~\bibnamefont
  {Huang}}, \bibinfo {author} {\bibfnamefont {H.}~\bibnamefont {Luo}}, \bibinfo
  {author} {\bibfnamefont {P.}~\bibnamefont {Gao}}, \bibinfo {author}
  {\bibfnamefont {M.}~\bibnamefont {García-Fernández}}, \bibinfo {author}
  {\bibfnamefont {L.}~\bibnamefont {Qiao}}, \ and\ \bibinfo {author}
  {\bibfnamefont {K.-J.}\ \bibnamefont {Zhou}},\ }\href {\doibase
  10.1038/s41563-022-01330-1} {\bibfield  {journal} {\bibinfo  {journal} {Nat.
  Mater.}\ }\textbf {\bibinfo {volume} {21}},\ \bibinfo {pages} {1116}
  (\bibinfo {year} {2022})}\BibitemShut {NoStop}%
\bibitem [{\citenamefont {Rossi}\ \emph {et~al.}(2022)\citenamefont {Rossi},
  \citenamefont {Osada}, \citenamefont {Choi}, \citenamefont {Agrestini},
  \citenamefont {Jost}, \citenamefont {Lee}, \citenamefont {Lu}, \citenamefont
  {Wang}, \citenamefont {Lee}, \citenamefont {Nag}, \citenamefont {Chuang},
  \citenamefont {Kuo}, \citenamefont {Lee}, \citenamefont {Moritz},
  \citenamefont {Devereaux}, \citenamefont {Shen}, \citenamefont {Lee},
  \citenamefont {Zhou}, \citenamefont {Hwang},\ and\ \citenamefont
  {Lee}}]{Rossi2022_CO_Reciprocal}%
  \BibitemOpen
  \bibfield  {author} {\bibinfo {author} {\bibfnamefont {M.}~\bibnamefont
  {Rossi}}, \bibinfo {author} {\bibfnamefont {M.}~\bibnamefont {Osada}},
  \bibinfo {author} {\bibfnamefont {J.}~\bibnamefont {Choi}}, \bibinfo {author}
  {\bibfnamefont {S.}~\bibnamefont {Agrestini}}, \bibinfo {author}
  {\bibfnamefont {D.}~\bibnamefont {Jost}}, \bibinfo {author} {\bibfnamefont
  {Y.}~\bibnamefont {Lee}}, \bibinfo {author} {\bibfnamefont {H.}~\bibnamefont
  {Lu}}, \bibinfo {author} {\bibfnamefont {B.~Y.}\ \bibnamefont {Wang}},
  \bibinfo {author} {\bibfnamefont {K.}~\bibnamefont {Lee}}, \bibinfo {author}
  {\bibfnamefont {A.}~\bibnamefont {Nag}}, \bibinfo {author} {\bibfnamefont
  {Y.-D.}\ \bibnamefont {Chuang}}, \bibinfo {author} {\bibfnamefont {C.-T.}\
  \bibnamefont {Kuo}}, \bibinfo {author} {\bibfnamefont {S.-J.}\ \bibnamefont
  {Lee}}, \bibinfo {author} {\bibfnamefont {B.}~\bibnamefont {Moritz}},
  \bibinfo {author} {\bibfnamefont {T.~P.}\ \bibnamefont {Devereaux}}, \bibinfo
  {author} {\bibfnamefont {Z.-X.}\ \bibnamefont {Shen}}, \bibinfo {author}
  {\bibfnamefont {J.-S.}\ \bibnamefont {Lee}}, \bibinfo {author} {\bibfnamefont
  {K.-J.}\ \bibnamefont {Zhou}}, \bibinfo {author} {\bibfnamefont {H.~Y.}\
  \bibnamefont {Hwang}}, \ and\ \bibinfo {author} {\bibfnamefont {W.-S.}\
  \bibnamefont {Lee}},\ }\href {\doibase 10.1038/s41567-022-01660-6} {\bibfield
   {journal} {\bibinfo  {journal} {Nat. Phys.}\ }\textbf {\bibinfo {volume}
  {18}},\ \bibinfo {pages} {869} (\bibinfo {year} {2022})}\BibitemShut
  {NoStop}%
\bibitem [{\citenamefont {Hepting}\ \emph {et~al.}(2020)\citenamefont
  {Hepting}, \citenamefont {Li}, \citenamefont {Jia}, \citenamefont {Lu},
  \citenamefont {Paris}, \citenamefont {Tseng}, \citenamefont {Feng},
  \citenamefont {Osada}, \citenamefont {Been}, \citenamefont {Hikita},
  \citenamefont {Chuang}, \citenamefont {Hussain}, \citenamefont {Zhou},
  \citenamefont {Nag}, \citenamefont {Garcia-Fernandez}, \citenamefont {Rossi},
  \citenamefont {Huang}, \citenamefont {Huang}, \citenamefont {Shen},
  \citenamefont {Schmitt}, \citenamefont {Hwang}, \citenamefont {Moritz},
  \citenamefont {Zaanen}, \citenamefont {Devereaux},\ and\ \citenamefont
  {Lee}}]{Hepting2020_XAS_Ni2p}%
  \BibitemOpen
  \bibfield  {author} {\bibinfo {author} {\bibfnamefont {M.}~\bibnamefont
  {Hepting}}, \bibinfo {author} {\bibfnamefont {D.}~\bibnamefont {Li}},
  \bibinfo {author} {\bibfnamefont {C.~J.}\ \bibnamefont {Jia}}, \bibinfo
  {author} {\bibfnamefont {H.}~\bibnamefont {Lu}}, \bibinfo {author}
  {\bibfnamefont {E.}~\bibnamefont {Paris}}, \bibinfo {author} {\bibfnamefont
  {Y.}~\bibnamefont {Tseng}}, \bibinfo {author} {\bibfnamefont
  {X.}~\bibnamefont {Feng}}, \bibinfo {author} {\bibfnamefont {M.}~\bibnamefont
  {Osada}}, \bibinfo {author} {\bibfnamefont {E.}~\bibnamefont {Been}},
  \bibinfo {author} {\bibfnamefont {Y.}~\bibnamefont {Hikita}}, \bibinfo
  {author} {\bibfnamefont {Y.-D.}\ \bibnamefont {Chuang}}, \bibinfo {author}
  {\bibfnamefont {Z.}~\bibnamefont {Hussain}}, \bibinfo {author} {\bibfnamefont
  {K.~J.}\ \bibnamefont {Zhou}}, \bibinfo {author} {\bibfnamefont
  {A.}~\bibnamefont {Nag}}, \bibinfo {author} {\bibfnamefont {M.}~\bibnamefont
  {Garcia-Fernandez}}, \bibinfo {author} {\bibfnamefont {M.}~\bibnamefont
  {Rossi}}, \bibinfo {author} {\bibfnamefont {H.~Y.}\ \bibnamefont {Huang}},
  \bibinfo {author} {\bibfnamefont {D.~J.}\ \bibnamefont {Huang}}, \bibinfo
  {author} {\bibfnamefont {Z.~X.}\ \bibnamefont {Shen}}, \bibinfo {author}
  {\bibfnamefont {T.}~\bibnamefont {Schmitt}}, \bibinfo {author} {\bibfnamefont
  {H.~Y.}\ \bibnamefont {Hwang}}, \bibinfo {author} {\bibfnamefont
  {B.}~\bibnamefont {Moritz}}, \bibinfo {author} {\bibfnamefont
  {J.}~\bibnamefont {Zaanen}}, \bibinfo {author} {\bibfnamefont {T.~P.}\
  \bibnamefont {Devereaux}}, \ and\ \bibinfo {author} {\bibfnamefont {W.~S.}\
  \bibnamefont {Lee}},\ }\href {\doibase 10.1038/s41563-019-0585-z} {\bibfield
  {journal} {\bibinfo  {journal} {Nat. Mater.}\ }\textbf {\bibinfo {volume}
  {19}},\ \bibinfo {pages} {381} (\bibinfo {year} {2020})}\BibitemShut
  {NoStop}%
\bibitem [{\citenamefont {Karp}\ \emph {et~al.}(2020)\citenamefont {Karp},
  \citenamefont {Botana}, \citenamefont {Norman}, \citenamefont {Park},
  \citenamefont {Zingl},\ and\ \citenamefont
  {Millis}}]{Karp2020_Chargehubbard}%
  \BibitemOpen
  \bibfield  {author} {\bibinfo {author} {\bibfnamefont {J.}~\bibnamefont
  {Karp}}, \bibinfo {author} {\bibfnamefont {A.~S.}\ \bibnamefont {Botana}},
  \bibinfo {author} {\bibfnamefont {M.~R.}\ \bibnamefont {Norman}}, \bibinfo
  {author} {\bibfnamefont {H.}~\bibnamefont {Park}}, \bibinfo {author}
  {\bibfnamefont {M.}~\bibnamefont {Zingl}}, \ and\ \bibinfo {author}
  {\bibfnamefont {A.}~\bibnamefont {Millis}},\ }\href {\doibase
  10.1103/PhysRevX.10.021061} {\bibfield  {journal} {\bibinfo  {journal} {Phys.
  Rev. X}\ }\textbf {\bibinfo {volume} {10}},\ \bibinfo {pages} {021061}
  (\bibinfo {year} {2020})}\BibitemShut {NoStop}%
\bibitem [{\citenamefont {Fowlie}\ \emph {et~al.}(2022)\citenamefont {Fowlie},
  \citenamefont {Hadjimichael}, \citenamefont {Martins}, \citenamefont {Li},
  \citenamefont {Osada}, \citenamefont {Wang}, \citenamefont {Lee},
  \citenamefont {Lee}, \citenamefont {Salman}, \citenamefont {Prokscha},
  \citenamefont {Triscone}, \citenamefont {Hwang},\ and\ \citenamefont
  {Suter}}]{Fowlie2022_Magnetic_NNO}%
  \BibitemOpen
  \bibfield  {author} {\bibinfo {author} {\bibfnamefont {J.}~\bibnamefont
  {Fowlie}}, \bibinfo {author} {\bibfnamefont {M.}~\bibnamefont
  {Hadjimichael}}, \bibinfo {author} {\bibfnamefont {M.~M.}\ \bibnamefont
  {Martins}}, \bibinfo {author} {\bibfnamefont {D.}~\bibnamefont {Li}},
  \bibinfo {author} {\bibfnamefont {M.}~\bibnamefont {Osada}}, \bibinfo
  {author} {\bibfnamefont {B.~Y.}\ \bibnamefont {Wang}}, \bibinfo {author}
  {\bibfnamefont {K.}~\bibnamefont {Lee}}, \bibinfo {author} {\bibfnamefont
  {Y.}~\bibnamefont {Lee}}, \bibinfo {author} {\bibfnamefont {Z.}~\bibnamefont
  {Salman}}, \bibinfo {author} {\bibfnamefont {T.}~\bibnamefont {Prokscha}},
  \bibinfo {author} {\bibfnamefont {J.-M.}\ \bibnamefont {Triscone}}, \bibinfo
  {author} {\bibfnamefont {H.~Y.}\ \bibnamefont {Hwang}}, \ and\ \bibinfo
  {author} {\bibfnamefont {A.}~\bibnamefont {Suter}},\ }\href {\doibase
  10.1038/s41567-022-01684-y} {\bibfield  {journal} {\bibinfo  {journal} {Nat.
  Phys.}\ }\textbf {\bibinfo {volume} {18}},\ \bibinfo {pages} {1043} (\bibinfo
  {year} {2022})}\BibitemShut {NoStop}%
\bibitem [{\citenamefont {Preziosi}\ \emph {et~al.}(2017)\citenamefont
  {Preziosi}, \citenamefont {Sander}, \citenamefont {Barthélémy},\ and\
  \citenamefont {Bibes}}]{Preziosi2017_Synthesis_NNO}%
  \BibitemOpen
  \bibfield  {author} {\bibinfo {author} {\bibfnamefont {D.}~\bibnamefont
  {Preziosi}}, \bibinfo {author} {\bibfnamefont {A.}~\bibnamefont {Sander}},
  \bibinfo {author} {\bibfnamefont {A.}~\bibnamefont {Barthélémy}}, \ and\
  \bibinfo {author} {\bibfnamefont {M.}~\bibnamefont {Bibes}},\ }\href
  {\doibase 10.1063/1.4975307} {\bibfield  {journal} {\bibinfo  {journal} {AIP
  Advances}\ }\textbf {\bibinfo {volume} {7}},\ \bibinfo {pages} {015210}
  (\bibinfo {year} {2017})}\BibitemShut {NoStop}%
\bibitem [{\citenamefont {Krieger}\ \emph
  {et~al.}(2022{\natexlab{b}})\citenamefont {Krieger}, \citenamefont {Raji},
  \citenamefont {Schlur}, \citenamefont {Versini}, \citenamefont {Bouillet},
  \citenamefont {Lenertz}, \citenamefont {Robert}, \citenamefont {Gloter},
  \citenamefont {Viart},\ and\ \citenamefont
  {Preziosi}}]{Krieger2022_synthesis_NNO}%
  \BibitemOpen
  \bibfield  {author} {\bibinfo {author} {\bibfnamefont {G.}~\bibnamefont
  {Krieger}}, \bibinfo {author} {\bibfnamefont {A.}~\bibnamefont {Raji}},
  \bibinfo {author} {\bibfnamefont {L.}~\bibnamefont {Schlur}}, \bibinfo
  {author} {\bibfnamefont {G.}~\bibnamefont {Versini}}, \bibinfo {author}
  {\bibfnamefont {C.}~\bibnamefont {Bouillet}}, \bibinfo {author}
  {\bibfnamefont {M.}~\bibnamefont {Lenertz}}, \bibinfo {author} {\bibfnamefont
  {J.}~\bibnamefont {Robert}}, \bibinfo {author} {\bibfnamefont
  {A.}~\bibnamefont {Gloter}}, \bibinfo {author} {\bibfnamefont
  {N.}~\bibnamefont {Viart}}, \ and\ \bibinfo {author} {\bibfnamefont
  {D.}~\bibnamefont {Preziosi}},\ }\href {\doibase 10.1088/1361-6463/aca54a}
  {\bibfield  {journal} {\bibinfo  {journal} {J. Phys. D: Appl. Phys.}\
  }\textbf {\bibinfo {volume} {56}},\ \bibinfo {pages} {024003} (\bibinfo
  {year} {2022}{\natexlab{b}})}\BibitemShut {NoStop}%
\bibitem [{\citenamefont {Fairley}\ \emph {et~al.}(2021)\citenamefont
  {Fairley}, \citenamefont {Fernandez}, \citenamefont {Richard‐Plouet},
  \citenamefont {Guillot-Deudon}, \citenamefont {Walton}, \citenamefont
  {Smith}, \citenamefont {Flahaut}, \citenamefont {Greiner}, \citenamefont
  {Biesinger}, \citenamefont {Tougaard}, \citenamefont {Morgan},\ and\
  \citenamefont {Baltrusaitis}}]{Fairley2021_CASA}%
  \BibitemOpen
  \bibfield  {author} {\bibinfo {author} {\bibfnamefont {N.}~\bibnamefont
  {Fairley}}, \bibinfo {author} {\bibfnamefont {V.}~\bibnamefont {Fernandez}},
  \bibinfo {author} {\bibfnamefont {M.}~\bibnamefont {Richard‐Plouet}},
  \bibinfo {author} {\bibfnamefont {C.}~\bibnamefont {Guillot-Deudon}},
  \bibinfo {author} {\bibfnamefont {J.}~\bibnamefont {Walton}}, \bibinfo
  {author} {\bibfnamefont {E.}~\bibnamefont {Smith}}, \bibinfo {author}
  {\bibfnamefont {D.}~\bibnamefont {Flahaut}}, \bibinfo {author} {\bibfnamefont
  {M.}~\bibnamefont {Greiner}}, \bibinfo {author} {\bibfnamefont
  {M.}~\bibnamefont {Biesinger}}, \bibinfo {author} {\bibfnamefont
  {S.}~\bibnamefont {Tougaard}}, \bibinfo {author} {\bibfnamefont
  {D.}~\bibnamefont {Morgan}}, \ and\ \bibinfo {author} {\bibfnamefont
  {J.}~\bibnamefont {Baltrusaitis}},\ }\href {\doibase
  https://doi.org/10.1016/j.apsadv.2021.100112} {\bibfield  {journal} {\bibinfo
   {journal} {Appl. Surf. Sci. Adv.}\ }\textbf {\bibinfo {volume} {5}},\
  \bibinfo {pages} {100112} (\bibinfo {year} {2021})}\BibitemShut {NoStop}%
\bibitem [{\citenamefont {van Veenendaal}\ and\ \citenamefont
  {Sawatzky}(1993)}]{Veenendaal1993_XPS_Ni2p}%
  \BibitemOpen
  \bibfield  {author} {\bibinfo {author} {\bibfnamefont {M.~A.}\ \bibnamefont
  {van Veenendaal}}\ and\ \bibinfo {author} {\bibfnamefont {G.~A.}\
  \bibnamefont {Sawatzky}},\ }\href {\doibase 10.1103/PhysRevLett.70.2459}
  {\bibfield  {journal} {\bibinfo  {journal} {Phys. Rev. Lett.}\ }\textbf
  {\bibinfo {volume} {70}},\ \bibinfo {pages} {2459} (\bibinfo {year}
  {1993})}\BibitemShut {NoStop}%
\bibitem [{\citenamefont {van
  Veenendaal}(2006)}]{Veenendaal2006_Ni2p_nonlocal}%
  \BibitemOpen
  \bibfield  {author} {\bibinfo {author} {\bibfnamefont {M.}~\bibnamefont {van
  Veenendaal}},\ }\href {\doibase 10.1103/PhysRevB.74.085118} {\bibfield
  {journal} {\bibinfo  {journal} {Phys. Rev. B}\ }\textbf {\bibinfo {volume}
  {74}},\ \bibinfo {pages} {085118} (\bibinfo {year} {2006})}\BibitemShut
  {NoStop}%
\bibitem [{\citenamefont {Fu}\ \emph {et~al.}(2020)\citenamefont {Fu},
  \citenamefont {Wang}, \citenamefont {Cheng}, \citenamefont {Pei},
  \citenamefont {Zhou}, \citenamefont {Chen}, \citenamefont {Wang},
  \citenamefont {Zhao}, \citenamefont {Jiang}, \citenamefont {Liu},
  \citenamefont {Huang}, \citenamefont {Wang}, \citenamefont {Zhao},
  \citenamefont {Yu}, \citenamefont {Ye}, \citenamefont {Wang},\ and\
  \citenamefont {Mei}}]{fu2020corelevel_XPS_NNO}%
  \BibitemOpen
  \bibfield  {author} {\bibinfo {author} {\bibfnamefont {Y.}~\bibnamefont
  {Fu}}, \bibinfo {author} {\bibfnamefont {L.}~\bibnamefont {Wang}}, \bibinfo
  {author} {\bibfnamefont {H.}~\bibnamefont {Cheng}}, \bibinfo {author}
  {\bibfnamefont {S.}~\bibnamefont {Pei}}, \bibinfo {author} {\bibfnamefont
  {X.}~\bibnamefont {Zhou}}, \bibinfo {author} {\bibfnamefont {J.}~\bibnamefont
  {Chen}}, \bibinfo {author} {\bibfnamefont {S.}~\bibnamefont {Wang}}, \bibinfo
  {author} {\bibfnamefont {R.}~\bibnamefont {Zhao}}, \bibinfo {author}
  {\bibfnamefont {W.}~\bibnamefont {Jiang}}, \bibinfo {author} {\bibfnamefont
  {C.}~\bibnamefont {Liu}}, \bibinfo {author} {\bibfnamefont {M.}~\bibnamefont
  {Huang}}, \bibinfo {author} {\bibfnamefont {X.}~\bibnamefont {Wang}},
  \bibinfo {author} {\bibfnamefont {Y.}~\bibnamefont {Zhao}}, \bibinfo {author}
  {\bibfnamefont {D.}~\bibnamefont {Yu}}, \bibinfo {author} {\bibfnamefont
  {F.}~\bibnamefont {Ye}}, \bibinfo {author} {\bibfnamefont {S.}~\bibnamefont
  {Wang}}, \ and\ \bibinfo {author} {\bibfnamefont {J.-W.}\ \bibnamefont
  {Mei}},\ }\href@noop {} {} (\bibinfo {year} {2020}),\ \Eprint
  {http://arxiv.org/abs/1911.03177} {arXiv:1911.03177 [cond-mat.supr-con]}
  \BibitemShut {NoStop}%
\bibitem [{\citenamefont {Bisogni}\ \emph {et~al.}(2016)\citenamefont
  {Bisogni}, \citenamefont {Catalano}, \citenamefont {Green}, \citenamefont
  {Gibert}, \citenamefont {Scherwitzl}, \citenamefont {Huang}, \citenamefont
  {Strocov}, \citenamefont {Zubko}, \citenamefont {Balandeh}, \citenamefont
  {Triscone}, \citenamefont {Sawatzky},\ and\ \citenamefont
  {Schmitt}}]{Bisogni2016_RIXS_CT}%
  \BibitemOpen
  \bibfield  {author} {\bibinfo {author} {\bibfnamefont {V.}~\bibnamefont
  {Bisogni}}, \bibinfo {author} {\bibfnamefont {S.}~\bibnamefont {Catalano}},
  \bibinfo {author} {\bibfnamefont {R.~J.}\ \bibnamefont {Green}}, \bibinfo
  {author} {\bibfnamefont {M.}~\bibnamefont {Gibert}}, \bibinfo {author}
  {\bibfnamefont {R.}~\bibnamefont {Scherwitzl}}, \bibinfo {author}
  {\bibfnamefont {Y.}~\bibnamefont {Huang}}, \bibinfo {author} {\bibfnamefont
  {V.~N.}\ \bibnamefont {Strocov}}, \bibinfo {author} {\bibfnamefont
  {P.}~\bibnamefont {Zubko}}, \bibinfo {author} {\bibfnamefont
  {S.}~\bibnamefont {Balandeh}}, \bibinfo {author} {\bibfnamefont {J.-M.}\
  \bibnamefont {Triscone}}, \bibinfo {author} {\bibfnamefont {G.}~\bibnamefont
  {Sawatzky}}, \ and\ \bibinfo {author} {\bibfnamefont {T.}~\bibnamefont
  {Schmitt}},\ }\href {\doibase 10.1038/ncomms13017} {\bibfield  {journal}
  {\bibinfo  {journal} {Nat. Commun.}\ }\textbf {\bibinfo {volume} {7}},\
  \bibinfo {pages} {13017} (\bibinfo {year} {2016})}\BibitemShut {NoStop}%
\bibitem [{\citenamefont {Higashi}\ \emph {et~al.}(2021)\citenamefont
  {Higashi}, \citenamefont {Winder}, \citenamefont {Kuneš},\ and\
  \citenamefont {Hariki}}]{Higashi2021_XPS_XAS_Ni2p}%
  \BibitemOpen
  \bibfield  {author} {\bibinfo {author} {\bibfnamefont {K.}~\bibnamefont
  {Higashi}}, \bibinfo {author} {\bibfnamefont {M.}~\bibnamefont {Winder}},
  \bibinfo {author} {\bibfnamefont {J.}~\bibnamefont {Kuneš}}, \ and\ \bibinfo
  {author} {\bibfnamefont {A.}~\bibnamefont {Hariki}},\ }\href {\doibase
  10.1103/PhysRevX.11.041009} {\bibfield  {journal} {\bibinfo  {journal} {Phys.
  Rev. X}\ }\textbf {\bibinfo {volume} {11}},\ \bibinfo {pages} {041009}
  (\bibinfo {year} {2021})}\BibitemShut {NoStop}%
\bibitem [{\citenamefont {Taguchi}\ \emph {et~al.}(2005)\citenamefont
  {Taguchi}, \citenamefont {Chainani}, \citenamefont {Horiba}, \citenamefont
  {Takata}, \citenamefont {Yabashi}, \citenamefont {Tamasaku}, \citenamefont
  {Nishino}, \citenamefont {Miwa}, \citenamefont {Ishikawa}, \citenamefont
  {Takeuchi}, \citenamefont {Yamamoto}, \citenamefont {Matsunami},
  \citenamefont {Shin}, \citenamefont {Yokoya}, \citenamefont {Ikenaga},
  \citenamefont {Kobayashi}, \citenamefont {Mochiku}, \citenamefont {Hirata},
  \citenamefont {Hori}, \citenamefont {Ishii}, \citenamefont {Nakamura},\ and\
  \citenamefont {Suzuki}}]{Taguchi2005_XPS_Cu_ZRS}%
  \BibitemOpen
  \bibfield  {author} {\bibinfo {author} {\bibfnamefont {M.}~\bibnamefont
  {Taguchi}}, \bibinfo {author} {\bibfnamefont {A.}~\bibnamefont {Chainani}},
  \bibinfo {author} {\bibfnamefont {K.}~\bibnamefont {Horiba}}, \bibinfo
  {author} {\bibfnamefont {Y.}~\bibnamefont {Takata}}, \bibinfo {author}
  {\bibfnamefont {M.}~\bibnamefont {Yabashi}}, \bibinfo {author} {\bibfnamefont
  {K.}~\bibnamefont {Tamasaku}}, \bibinfo {author} {\bibfnamefont
  {Y.}~\bibnamefont {Nishino}}, \bibinfo {author} {\bibfnamefont
  {D.}~\bibnamefont {Miwa}}, \bibinfo {author} {\bibfnamefont {T.}~\bibnamefont
  {Ishikawa}}, \bibinfo {author} {\bibfnamefont {T.}~\bibnamefont {Takeuchi}},
  \bibinfo {author} {\bibfnamefont {K.}~\bibnamefont {Yamamoto}}, \bibinfo
  {author} {\bibfnamefont {M.}~\bibnamefont {Matsunami}}, \bibinfo {author}
  {\bibfnamefont {S.}~\bibnamefont {Shin}}, \bibinfo {author} {\bibfnamefont
  {T.}~\bibnamefont {Yokoya}}, \bibinfo {author} {\bibfnamefont
  {E.}~\bibnamefont {Ikenaga}}, \bibinfo {author} {\bibfnamefont
  {K.}~\bibnamefont {Kobayashi}}, \bibinfo {author} {\bibfnamefont
  {T.}~\bibnamefont {Mochiku}}, \bibinfo {author} {\bibfnamefont
  {K.}~\bibnamefont {Hirata}}, \bibinfo {author} {\bibfnamefont
  {J.}~\bibnamefont {Hori}}, \bibinfo {author} {\bibfnamefont {K.}~\bibnamefont
  {Ishii}}, \bibinfo {author} {\bibfnamefont {F.}~\bibnamefont {Nakamura}}, \
  and\ \bibinfo {author} {\bibfnamefont {T.}~\bibnamefont {Suzuki}},\ }\href
  {\doibase 10.1103/PhysRevLett.95.177002} {\bibfield  {journal} {\bibinfo
  {journal} {Phys. Rev. Lett.}\ }\textbf {\bibinfo {volume} {95}},\ \bibinfo
  {pages} {177002} (\bibinfo {year} {2005})}\BibitemShut {NoStop}%
\bibitem [{\citenamefont {van Veenendaal}\ and\ \citenamefont
  {Sawatzky}(1994)}]{Veenendaal1994_XPS_Cu_Theory}%
  \BibitemOpen
  \bibfield  {author} {\bibinfo {author} {\bibfnamefont {M.~A.}\ \bibnamefont
  {van Veenendaal}}\ and\ \bibinfo {author} {\bibfnamefont {G.~A.}\
  \bibnamefont {Sawatzky}},\ }\href {\doibase 10.1103/PhysRevB.49.3473}
  {\bibfield  {journal} {\bibinfo  {journal} {Phys. Rev. B}\ }\textbf {\bibinfo
  {volume} {49}},\ \bibinfo {pages} {3473} (\bibinfo {year}
  {1994})}\BibitemShut {NoStop}%
\bibitem [{\citenamefont {Goodge}\ \emph {et~al.}(2023)\citenamefont {Goodge},
  \citenamefont {Geisler}, \citenamefont {Lee}, \citenamefont {Osada},
  \citenamefont {Wang}, \citenamefont {Li}, \citenamefont {Hwang},
  \citenamefont {Pentcheva},\ and\ \citenamefont
  {Kourkoutis}}]{Goodge2023_NNO_Interface_TEM}%
  \BibitemOpen
  \bibfield  {author} {\bibinfo {author} {\bibfnamefont {B.~H.}\ \bibnamefont
  {Goodge}}, \bibinfo {author} {\bibfnamefont {B.}~\bibnamefont {Geisler}},
  \bibinfo {author} {\bibfnamefont {K.}~\bibnamefont {Lee}}, \bibinfo {author}
  {\bibfnamefont {M.}~\bibnamefont {Osada}}, \bibinfo {author} {\bibfnamefont
  {B.~Y.}\ \bibnamefont {Wang}}, \bibinfo {author} {\bibfnamefont
  {D.}~\bibnamefont {Li}}, \bibinfo {author} {\bibfnamefont {H.~Y.}\
  \bibnamefont {Hwang}}, \bibinfo {author} {\bibfnamefont {R.}~\bibnamefont
  {Pentcheva}}, \ and\ \bibinfo {author} {\bibfnamefont {L.~F.}\ \bibnamefont
  {Kourkoutis}},\ }\href {\doibase 10.1038/s41563-023-01510-7} {\bibfield
  {journal} {\bibinfo  {journal} {Nat. Mater.}\ }\textbf {\bibinfo {volume}
  {22}},\ \bibinfo {pages} {466} (\bibinfo {year} {2023})}\BibitemShut
  {NoStop}%
\bibitem [{\citenamefont {Gu}\ \emph {et~al.}(2020)\citenamefont {Gu},
  \citenamefont {Li}, \citenamefont {Wan}, \citenamefont {Li}, \citenamefont
  {Guo}, \citenamefont {Yang}, \citenamefont {Li}, \citenamefont {Zhu},
  \citenamefont {Pan}, \citenamefont {Nie},\ and\ \citenamefont
  {Wen}}]{Gu2020_STM_NNSO}%
  \BibitemOpen
  \bibfield  {author} {\bibinfo {author} {\bibfnamefont {Q.}~\bibnamefont
  {Gu}}, \bibinfo {author} {\bibfnamefont {Y.}~\bibnamefont {Li}}, \bibinfo
  {author} {\bibfnamefont {S.}~\bibnamefont {Wan}}, \bibinfo {author}
  {\bibfnamefont {H.}~\bibnamefont {Li}}, \bibinfo {author} {\bibfnamefont
  {W.}~\bibnamefont {Guo}}, \bibinfo {author} {\bibfnamefont {H.}~\bibnamefont
  {Yang}}, \bibinfo {author} {\bibfnamefont {Q.}~\bibnamefont {Li}}, \bibinfo
  {author} {\bibfnamefont {X.}~\bibnamefont {Zhu}}, \bibinfo {author}
  {\bibfnamefont {X.}~\bibnamefont {Pan}}, \bibinfo {author} {\bibfnamefont
  {Y.}~\bibnamefont {Nie}}, \ and\ \bibinfo {author} {\bibfnamefont {H.-H.}\
  \bibnamefont {Wen}},\ }\href {\doibase 10.1038/s41467-020-19908-1} {\bibfield
   {journal} {\bibinfo  {journal} {Nat. Commun.}\ }\textbf {\bibinfo {volume}
  {11}},\ \bibinfo {pages} {6027} (\bibinfo {year} {2020})}\BibitemShut
  {NoStop}%
\bibitem [{\citenamefont {Chen}\ \emph {et~al.}(2022)\citenamefont {Chen},
  \citenamefont {Osada}, \citenamefont {Li}, \citenamefont {Been},
  \citenamefont {Chen}, \citenamefont {Hashimoto}, \citenamefont {Lu},
  \citenamefont {Mo}, \citenamefont {Lee}, \citenamefont {Wang}, \citenamefont
  {Rodolakis}, \citenamefont {McChesney}, \citenamefont {Jia}, \citenamefont
  {Moritz}, \citenamefont {Devereaux}, \citenamefont {Hwang},\ and\
  \citenamefont {Shen}}]{CHEN20221806_XPS_VB_Ni3d}%
  \BibitemOpen
  \bibfield  {author} {\bibinfo {author} {\bibfnamefont {Z.}~\bibnamefont
  {Chen}}, \bibinfo {author} {\bibfnamefont {M.}~\bibnamefont {Osada}},
  \bibinfo {author} {\bibfnamefont {D.}~\bibnamefont {Li}}, \bibinfo {author}
  {\bibfnamefont {E.~M.}\ \bibnamefont {Been}}, \bibinfo {author}
  {\bibfnamefont {S.-D.}\ \bibnamefont {Chen}}, \bibinfo {author}
  {\bibfnamefont {M.}~\bibnamefont {Hashimoto}}, \bibinfo {author}
  {\bibfnamefont {D.}~\bibnamefont {Lu}}, \bibinfo {author} {\bibfnamefont
  {S.-K.}\ \bibnamefont {Mo}}, \bibinfo {author} {\bibfnamefont
  {K.}~\bibnamefont {Lee}}, \bibinfo {author} {\bibfnamefont {B.~Y.}\
  \bibnamefont {Wang}}, \bibinfo {author} {\bibfnamefont {F.}~\bibnamefont
  {Rodolakis}}, \bibinfo {author} {\bibfnamefont {J.~L.}\ \bibnamefont
  {McChesney}}, \bibinfo {author} {\bibfnamefont {C.}~\bibnamefont {Jia}},
  \bibinfo {author} {\bibfnamefont {B.}~\bibnamefont {Moritz}}, \bibinfo
  {author} {\bibfnamefont {T.~P.}\ \bibnamefont {Devereaux}}, \bibinfo {author}
  {\bibfnamefont {H.~Y.}\ \bibnamefont {Hwang}}, \ and\ \bibinfo {author}
  {\bibfnamefont {Z.-X.}\ \bibnamefont {Shen}},\ }\href {\doibase
  https://doi.org/10.1016/j.matt.2022.01.020} {\bibfield  {journal} {\bibinfo
  {journal} {Matter}\ }\textbf {\bibinfo {volume} {5}},\ \bibinfo {pages}
  {1806} (\bibinfo {year} {2022})}\BibitemShut {NoStop}%
\end{thebibliography}

%

\end{document}